\documentclass{article}

\usepackage{IEEEtrantools}
\usepackage{spconf}
\usepackage{microtype}
\usepackage{cite}
\usepackage[pdftex]{graphicx}
\usepackage{amsmath, amssymb}
\usepackage{algorithmicx}
\usepackage{array}
\usepackage[caption=false,font=footnotesize,subrefformat=parens,labelformat=parens]{subfig}
\usepackage{url}
\usepackage{booktabs}
\usepackage{tikz}

\DeclareMathOperator{\cov}{cov}
\DeclareMathOperator{\tr}{tr}
\DeclareMathOperator*{\argmax}{arg\,max}

\title{Learning While Navigating: A Practical System Based on Variational Gaussian Process State-Space Model and Smartphone Sensory Data}
\name{Ang Xie$^{\star \dagger}$ \qquad Feng Yin$^{\dagger}$ \qquad Bo Ai$^{\star}$ \qquad Shuguang Cui$^{\dagger}$\thanks{Correspondence author: Feng Yin (Email: yinfeng@cuhk.edu.cn).}}
\address{$^{\star}$ State Key Laboratory of Rail Traffic Control and Safety, Beijing Jiaotong University, China\\ $^{\dagger}$ SRIBD and The Chinese University of Hong Kong, Shenzhen, China}
\begin{document}
\topmargin=0mm
\ninept
\maketitle
\begin{abstract}
We implement a wireless indoor navigation system based on the variational Gaussian process state-space model (GPSSM) with smartphone-collected WiFi received signal strength (RSS) and inertial measurement unit (IMU) readings. We adapt the existing variational GPSSM framework to wireless navigation scenarios, and provide a practical learning procedure for the variational GPSSM. The proposed system explores both the expressive power of the non-parametric Gaussian process model and its natural mechanism for integrating the state-of-the-art navigation techniques designed upon state-space model. Experimental results obtained from a real office environment validate the outstanding performance of the variational GPSSM in comparison with the traditional parametric state-space model in terms of navigation accuracy.
\end{abstract}
\begin{keywords}
Indoor navigation, Gaussian process, received signal strength, state-space model, smartphone sensory data.
\end{keywords}
\section{Introduction and Related Works}
Indoor navigation is typically built upon state-space model (SSM) with an evolving latent state, which comprises position, velocity, and acceleration \cite{Gustafsson10}. The SSM fuses position related measurements from different sources, including the inertial measurement unit (IMU) that measures the inertial motion, the WiFi access point (AP) that measures the received signal strength (RSS), and the Bluetooth beacons that measures proximity, etc. For traditional parametric SSMs, inference on the state can be done via Bayesian filtering algorithms, such as the Kalman filter, the extended Kalman filter, the unscented Kalman filter \cite{Julier04}, and the particle filter \cite{Doucet09}. However, parametric SSMs are lack of the ability to exploit useful motion patterns from the data, and moreover they are hard to specify when the underlying dynamics is not well understood \cite{Frigola15}.

With the explosion of data and the ever-increasing computational power, data-driven models and algorithms become popular. In this paper, we focus on the application of Gaussian process (GP) regression models, which constitute a class of important Bayesian non-parametric models for machine learning \cite{RW06}. Due to their outstanding performance in function approximation with a natural uncertainty bound, GPs have been adopted to approximate the nonlinear functions in SSMs, leading to the promising Gaussian process state-space model (GPSSM) \cite{Frigola15}. Early variants of the GPSSM were learned by finding the maximum a posteriori (MAP) estimates of the latent states, generating various successful positioning applications, among others the RSS-based WiFi localization \cite{Ferris07}, the human motion capture \cite{Wang08}, and the IMU-based slotcar tracking \cite{Ko11}, etc. The first fully probabilistic learning procedure of the GPSSM was proposed in \cite{Frigola13} using particle Markov Chain Monte Carlo (PMCMC). In order to reduce the heavy computational load of the sampling method used in \cite{Frigola13}, a number of different variational learning procedures were developed in \cite{Frigola14a, Eleftheriadis17, Ialongo18a, Ialongo18b} upon the classical variational sparse GP framework \cite{Titsias09}. Due to the implementation difficulties, their performance in real-world applications remains largely unexamined. 

The contributions of this work are summarized as follows. We propose a practical learning procedure for the variational GPSSM \cite{Frigola15} in the context of indoor navigation using smartphone sensory data. %The selection of inducing inputs of the variation sparse GP framework is achieved by using the smoothing distribution of representative trajectories.
The proposed learning procedure processes the measurement function and transition function in the GPSSM in a sequential manner and reduces the complexity by performing optimization on separate groups of the model hyperparameters. To maintain a competent navigation accuracy, we also integrate a cheap RSS-based WiFi localization technique and the pedestrian dead reckoning (PDR) approach, which leverages smartphone built-in IMU to perform step detection and walking direction estimation. The implemented GPSSM learning procedure is validated with real RSS measurements and IMU readings collected in an office environment. Comparisons with the classical parametric SSM in terms of navigation accuracy are provided.

The rest of this paper is organized as follows. Section~\ref{sec:GPSSM} gives a brief introduction to the recently developed GPSSM \cite{Frigola15}. Section~\ref{sec:Variational Learning} presents our proposed navigation system based on the data-driven, non-parametric variational GPSSM. Section~\ref{sec:Experiments} provides experimental results that confirm the superior performance of the proposed system. Lastly, Section~\ref{sec:Conclusion} concludes this paper.

\vspace{-0.4\baselineskip}
\section{Gaussian Process State-Space Models}
\label{sec:GPSSM}
State space models are suitable for modeling a measured time series $\boldsymbol{y}_{1:T}\triangleq\{\boldsymbol{y}_t\}_{t=1}^T$ with latent states $\boldsymbol{x}_{0:T}\triangleq\{\boldsymbol{x}_t\}_{t=0}^T$. A SSM is defined by a transition function, $f:\mathbb{R}^{D_x}\times\mathbb{R}^{D_u}\rightarrow\mathbb{R}^{D_x}$, and a measurement function, $g:\mathbb{R}^{D_x}\rightarrow\mathbb{R}^{D_y}$, as
\begin{IEEEeqnarray}{rCl}
\boldsymbol{x}_t & = & f(\boldsymbol{x}_{t-1}, \boldsymbol{u}_t) + \boldsymbol{q}_{t-1},\IEEEyesnumber\IEEEyessubnumber\\
\boldsymbol{y}_t & = & g(\boldsymbol{x}_t) + \boldsymbol{r}_t,\IEEEyessubnumber
\end{IEEEeqnarray}
where $\boldsymbol{x}_t \in \mathbb{R}^{D_x}$ is the latent state, $\boldsymbol{y}_t \in \mathbb{R}^{D_y}$ is the measurement, $\boldsymbol{u}_t \in \mathbb{R}^{D_u}$ is the control input, $\boldsymbol{q}_{t-1}$ and $\boldsymbol{r}_t$ are the process noise and measurement noise, respectively.

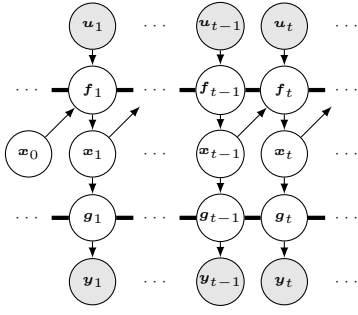
\begin{figure}[t]
	\centering
	\tiny
	\begin{tikzpicture}[align = center, latent/.style={circle, draw, text width = 0.45cm}, observed/.style={circle, draw, fill=gray!20, text width = 0.45cm}, transparent/.style={circle, text width = 0.45cm}, node distance=0.85cm]
	\node[latent](x0) {$\boldsymbol{x}_0$};
	\node[latent, right of=x0](x1) {$\boldsymbol{x}_{1}$};
	\node[transparent, right of=x1](x2) {$\cdots$};
	\node[latent, right of=x2](xt-1) {$\!\!\boldsymbol{x}_{t-1}\!\!$};
	\node[latent, right of=xt-1](xt) {$\boldsymbol{x}_{t}$};
	\node[transparent, right of=xt](xinf) {$\cdots$};
	\node[transparent, above of=x0](f0) {$\cdots$};
	\node[latent, above of=x1](f1) {$\boldsymbol{f}_{1}$};
	\node[transparent, right of=f1](f2) {$\cdots$};
	\node[latent, above of=xt-1](ft-1) {$\!\!\boldsymbol{f}_{t-1}\!\!$};
	\node[latent, above of=xt](ft) {$\boldsymbol{f}_{t}$};
	\node[transparent, right of=ft](finf) {$\cdots$};
	\node[observed, above of=f1](u1) {$\boldsymbol{u}_{1}$};
	\node[transparent, right of=u1](u2) {$\cdots$};
	\node[observed, above of=ft-1](ut-1) {$\!\!\boldsymbol{u}_{t-1}\!\!$};
	\node[observed, above of=ft](ut) {$\boldsymbol{u}_{t}$};
	\node[transparent, right of=ut](uinf) {$\cdots$};
	\node[transparent, below of=x0](g0) {$\cdots$};
	\node[latent, below of=x1](g1) {$\boldsymbol{g}_{1}$};
	\node[transparent, below of=x2](g2) {$\cdots$};
	\node[latent, below of=xt-1](gt-1) {$\!\!\boldsymbol{g}_{t-1}\!\!$};
	\node[latent, right of=gt-1](gt) {$\boldsymbol{g}_{t}$};
	\node[transparent, right of=gt](ginf) {$\cdots$};
	\node[observed, below of=g1](y1) {$\boldsymbol{y}_{1}$};
	\node[transparent, below of=g2](y2) {$\cdots$};
	\node[observed, below of=gt-1](yt-1) {$\!\!\boldsymbol{y}_{t-1}\!\!$};
	\node[observed, right of=yt-1](yt) {$\boldsymbol{y}_{t}$};
	\node[transparent, right of=yt](yinf) {$\cdots$};
	\draw[-latex] (x0) -- (f1);
	\draw[-latex] (f1) -- (x1);
	\draw[-latex] (x1) -- (f2);
	\draw[-latex] (ft-1) -- (xt-1);
	\draw[-latex] (xt-1) -- (ft);
	\draw[-latex] (ft) -- (xt);
	\draw[-latex] (xt) -- (finf);
	\draw[-latex] (u1) -- (f1);
	\draw[-latex] (ut-1) -- (ft-1);
	\draw[-latex] (ut) -- (ft);
	\draw[-latex] (x1) -- (g1);
	\draw[-latex] (xt-1) -- (gt-1);
	\draw[-latex] (xt) -- (gt);
	\draw[-latex] (g1) -- (y1);
	\draw[-latex] (gt-1) -- (yt-1);
	\draw[-latex] (gt) -- (yt);
	\draw[ultra thick]
	(f0) -- (f1)
	(f1) -- (f2)
	(f2) -- (ft-1)
	(ft-1) -- (ft)
	(ft) -- (finf)
	(g0) -- (g1)
	(g1) -- (g2)
	(g2) -- (gt-1)
	(gt-1) -- (gt)
	(gt) -- (ginf);
	\end{tikzpicture}
	\caption{Graphical model of GPSSM. The shaded nodes denote measurements while the transparent nodes denote latent variables.}
	\label{fig:graphical model}
	\vspace{-1\baselineskip}
\end{figure}

Traditional SSMs restrict $f$ and $g$ to be parametric functions, whose parameters can be learned through the expectation-maximization (EM) \cite{Schon11} or Markov chain Monte Carlo (MCMC) \cite{Andrieu10} methods. The GPSSMs \cite{Frigola13, Frigola14a, Frigola14b, Eleftheriadis17, Ialongo18a, Ialongo18b} form a popular class of non-parametric SSMs, which can handle big data without model saturation \cite{Frigola15}. In a GPSSM, both the transition function $f$ and the measurement function $g$ are modeled by GPs that are completely specified by their mean function $m(\boldsymbol{\cdot})$ and kernel function (a.k.a.\ covariance function) $k(\boldsymbol{\cdot}, \boldsymbol{\cdot})$. According to \cite{Frigola15}, the generative model for a GPSSM is given by
\begin{IEEEeqnarray}{rCl}
f(\boldsymbol{x}, \boldsymbol{u}) & \sim & \mathcal{GP}(m_f(\boldsymbol{x}, \boldsymbol{u}), k_f((\boldsymbol{x}, \boldsymbol{u}), (\boldsymbol{x}', \boldsymbol{u}'))),\IEEEyesnumber\IEEEyessubnumber\\
g(\boldsymbol{x}) & \sim & \mathcal{GP}(m_g(\boldsymbol{x}), k_g(\boldsymbol{x}, \boldsymbol{x}')),\IEEEyessubnumber\\
%\boldsymbol{x}_0 & \sim & p(\boldsymbol{x}_0),\IEEEyessubnumber\\
\boldsymbol{x}_t | \boldsymbol{f}_t & \sim & \mathcal{N}(\boldsymbol{x}_t | \boldsymbol{f}_t, \boldsymbol{Q}), \text{where } \boldsymbol{f}_t \triangleq  f(\boldsymbol{x}_{t-1}, \boldsymbol{u}_t),\IEEEyessubnumber\\
\boldsymbol{y}_t | \boldsymbol{g}_t & \sim & \mathcal{N}(\boldsymbol{y}_t | \boldsymbol{g}_t, \boldsymbol{R}), \text{where } \boldsymbol{g}_t \triangleq g(\boldsymbol{x}_t),\IEEEyessubnumber
\end{IEEEeqnarray}
with the model hyperparameters $\{\boldsymbol{\theta}_f, \boldsymbol{\theta}_g, \boldsymbol{Q}, \boldsymbol{R}\}$, where $\boldsymbol{\theta}_f$ and $\boldsymbol{\theta}_g$ are specially known as the kernel hyperparameters of the GPs, $\boldsymbol{Q}$ and $\boldsymbol{R}$ are the covariance matrices of the Gaussian process noise and the measurement noise, respectively. The initial state follows certain distribution $\boldsymbol{x}_0 \sim p(\boldsymbol{x}_0)$. For clarity, Fig.~\ref{fig:graphical model} shows the graphical model of a GPSSM. 
%Hyperparameter learning in a GPSSM is particularly challenging due to the linking between the consecutive latent states and the uncertainties added to them. 
%Unfortunately, the double GP structure of GPSSM suffers from non-identifiability issues \cite{Frigola15}, impeding its way towards practical usage.
%which can be eliminated to some extent by converting the model to a single GP structure, i.e., allowing a linear measurement function $g$ through augmenting the state-space to incorporate the non-linear measurement function.

\vspace{-0.2\baselineskip}
\section{Proposed Indoor Navigation System}
\label{sec:Variational Learning}
In this section, we combine the cutting-edge GPSSM with the state-of-the-art techniques for indoor navigation, leading to a practical learning procedure based on WiFi RSS measurements and smartphone IMU readings. The proposed learning procedure reduces the optimization complexity by handling the measurement function and transition function in a sequential manner, thereby alleviating non-identifiability issues.

%\vspace{-0.4\baselineskip}
\subsection{Learning of Measurement Function $\boldsymbol{g}$}

\subsubsection{WiFi RSS-based Localization as Pretraining}
Since GPS has poor coverage for indoor environments, most of the recent research efforts on indoor localization have been made to obtain position estimates from local wireless systems, such as WiFi, Bluetooth, RFID, etc. In this paper, we focus on the WiFi system and use RSS as position related measurement, which is cheap and easy to capture on smartphones\cite{Yin17b}. However, it is also well known that RSS measurements can be too noisy to be used for accurate localization\cite{Yin13}. But as we will see later, the powerful variational GPSSM could largely remedy this drawback of RSS.

To describe the signal propagation, the classic linear log-distance model for the RSS measurements generated by a single WiFi AP is given by 
%
%\begin{equation}
%r_i = \underbrace{A + 10B\log_{10}\frac{d_i}{d_0}}_{\mu_i} + e_i, \quad i = 1,2,\ldots, M_\text{RSS},
%\end{equation}
%
$r_i = A + 10B\log_{10}(d_i/d_0) + e_i, i = 1,2,\ldots, M_\text{RSS}$,
where $r_i$ is the $i$-th RSS collected through multiple scans, $A$ and $B$ are the path loss parameters, $d_0$ is the reference distance, and $d_i \triangleq \|\boldsymbol{p}_i - \boldsymbol{p}_\text{AP}\|$ is the Euclidean distance between the $3$-D AP position $\boldsymbol{p}_\text{AP}$ and the smartphone's $3$-D position $\boldsymbol{p}_i \triangleq [\boldsymbol{x}_i^T, h_i]^T$ where the RSS is scanned. For simplicity, we assume that the height $h_i$ of smartphone above ground is fixed throughout this paper, i.e., we hold the smartphone at a constant level. We assume the noise terms, $e_i$, to be Gaussian i.i.d.\ with zero mean and variance $\sigma^2$. The parameters $A$, $B$, and $\sigma$ for each AP are estimated via linear least-squares (LLS) fitting.

After having obtained the propagation model parameters of all APs, we then perform WiFi localization based on the maximum likelihood (ML) estimation. Given the collected RSS measurements through a single scan at an arbitrary position, the smartphone's $2$-D position $\boldsymbol{x}$ can be estimated through maximizing the log-likelihood function, namely, $\boldsymbol{y} = \argmax_{\boldsymbol{x}} l(\boldsymbol{x})$, like in \cite{Yin15}.
%various log-likelihood functions $l(\boldsymbol{x})$ can be formulated depending on precision requirements \cite{Yin15}.
%Assume that we collect $M_\text{AP}$ RSS measurements through a single scan at an arbitrary position, each belonging to a nearby AP. The corresponding log-likelihood function can be formulated, according to \cite{Yin15}, as
%
%\begin{equation}\label{eq:truncation log-likelihood}
%l(\boldsymbol{x}) \triangleq \sum_{j=1}^{M_\text{AP}}\log \frac{\frac{1}{\sqrt{2\pi\sigma^2_j}}\exp\left[-\frac{(r_j-\mu_j(\boldsymbol{x}))^2}{2\sigma_j^2}\right]}{\frac{1}{2}\Big[1-\erf(\frac{P_\text{dec}-\mu_j(\boldsymbol{x})}{\sqrt{2\sigma_j^2}})\Big]},
%\end{equation}
%
%where $\erf(\cdot)$ stands for the standard Gaussian error function, and $P_\text{dec}$ is the RSS truncation threshold. 

%
%\begin{equation}
%\boldsymbol{y} = \argmax_{\boldsymbol{x}} l(\boldsymbol{x}).
%\end{equation}

\vspace{-0.2\baselineskip}
\subsubsection{GP Model for Refinement}
\label{subsubsec:GP Optimization}
WiFi RSS-based localization can only provide a coarse position estimate $\boldsymbol{y}$ of the latent state, i.e., a $2$-D position $\boldsymbol{x}$. The quality of the estimate depends on the position to be determined, the distribution of the APs, and how complex the underlying radio propagation condition is. To better reflect complex indoor environment,  a nonlinear GP model was introduced in \cite{Yin15}, which represents a RSS value in terms of the $2$-D position. However, their approach requires each AP to train an individual GP model, causing a huge computational burden. For simplicity, we represent the position estimates in terms of the $2$-D position through a single GP model $\boldsymbol{y} = g(\boldsymbol{x}) + \boldsymbol{r}$, where $g(\boldsymbol{x})$ is a GP with the kernel hyperparameters $\boldsymbol{\theta}_g$, and $\boldsymbol{r}$ is a vector of measurement noise terms that follows $\boldsymbol{r} \sim \mathcal{N}(\boldsymbol{0}, \boldsymbol{R})$. We use a linear mean function $m_g(\boldsymbol{x}) = \boldsymbol{x}$ in the GP model. Given the training dataset $\boldsymbol{x}_{1:N} \triangleq \{\boldsymbol{x}_n\}_{n=1}^N$, $\boldsymbol{y}_{1:N} \triangleq \{\boldsymbol{y}_n\}_{n=1}^N$, we write the log-marginal likelihood function of the position estimates as
\begin{IEEEeqnarray}{l}\label{eq:GP log-marginal likelihood}
\log p(\boldsymbol{y}_{1:N}|\boldsymbol{x}_{1:N}, \boldsymbol{\theta}_g, \boldsymbol{R})\nonumber\\
= \log \mathcal{N}(\boldsymbol{y}_{1:N}|\boldsymbol{m}_g(\boldsymbol{x}_{1:N}), \boldsymbol{K}_g(\boldsymbol{x}_{1:N}, \boldsymbol{x}_{1:N}) + \boldsymbol{I}_N \otimes \boldsymbol{R}),\IEEEeqnarraynumspace
\end{IEEEeqnarray}
where $\boldsymbol{m}_g(\boldsymbol{x}_{1:N}) \triangleq [m_g(\boldsymbol{x}_1)^T, m_g(\boldsymbol{x}_2)^T, \ldots, m_g(\boldsymbol{x}_N)^T]^T$, and $\boldsymbol{K}_g(\boldsymbol{x}_{1:N}, \boldsymbol{x}_{1:N})$ is the shorthand notation of the kernel matrix, whose element in the $i, j$ position is $k_g(\boldsymbol{x}_i, \boldsymbol{x}_j)$.
%
%\begin{IEEEeqnarray}{l}
%\boldsymbol{K}_g(\boldsymbol{x}_{1:N}, \boldsymbol{x}_{1:N})\nonumber\\ \triangleq
%\begin{bmatrix}
%k_g(\boldsymbol{x}_1, \boldsymbol{x}_1) & k_g(\boldsymbol{x}_1, \boldsymbol{x}_2) & \ldots & k_g(\boldsymbol{x}_1, \boldsymbol{x}_N)\\
%k_g(\boldsymbol{x}_2, \boldsymbol{x}_1) & k_g(\boldsymbol{x}_2, \boldsymbol{x}_2) & \ldots & k_g(\boldsymbol{x}_2, \boldsymbol{x}_N)\\
%\vdots & \vdots & \ddots & \vdots \\
%k_g(\boldsymbol{x}_N, \boldsymbol{x}_1) & k_g(\boldsymbol{x}_N, \boldsymbol{x}_2) & \ldots & k_g(\boldsymbol{x}_N, \boldsymbol{x}_N)
%\end{bmatrix}.
%\end{IEEEeqnarray}
%
Note that in practice the mean function $m_g(\boldsymbol{\cdot})$ and the kernel function $k_g(\boldsymbol{\cdot}, \boldsymbol{\cdot})$ are in forms of a vector and a matrix of the same size as the state, respectively, analogous to the notations used for multi-output GPs \cite{Frigola15}. We optimize $\boldsymbol{\theta}_g$ and $\boldsymbol{R}$ through maximizing the log-marginal likelihood function in \eqref{eq:GP log-marginal likelihood}. Conventional optimization routines include the conjugate gradient (CG) and the limited-memory BFGS (L-BFGS) \cite{Rasmussen10}, advanced ones can be obtained based on the alternating direction method of multipliers (ADMM) \cite{Xie19, Xu19}. Once the GP hyperparameters have been optimized, the posterior distribution for a test input $\boldsymbol{x}_*$, given the training dataset, can be derived in an analytical Gaussian form. Concretely, this posterior distribution is given by
$p(\boldsymbol{y}_*|\boldsymbol{x}_{1:N}, \boldsymbol{y}_{1:N}, \boldsymbol{x}_*) \sim \mathcal{N}(\bar{\boldsymbol{y}}_*, \cov(\boldsymbol{y}_*))$, where $\bar{\boldsymbol{y}}_*  = m_g(\boldsymbol{x}_*) + \boldsymbol{K}_g (\boldsymbol{x}_*, \boldsymbol{x}_{1:N})[\boldsymbol{K}_g(\boldsymbol{x}_{1:N}, \boldsymbol{x}_{1:N}) + \boldsymbol{I}_N \otimes \boldsymbol{R}]^{-1}[\boldsymbol{y}_{1:N} - \boldsymbol{m}_g (\boldsymbol{x}_{1:N})]$, and $\cov(\boldsymbol{y}_*) =  k_g(\boldsymbol{x}_*, \boldsymbol{x}_*) - \boldsymbol{K}_g(\boldsymbol{x}_*, \boldsymbol{x}_{1:N})[\boldsymbol{K}_g(\boldsymbol{x}_{1:N}, \boldsymbol{x}_{1:N}) + \boldsymbol{I}_N  \otimes \boldsymbol{R}]^{-1}\boldsymbol{K}_g(\boldsymbol{x}_{1:N}, \boldsymbol{x}_*)$. For deriving this posterior distribution, interested readers can refer \cite{RW06} for more details. 
%
%\begin{IEEEeqnarray}{rCl}\label{eq:GP posterior}
%\bar{\boldsymbol{y}}_*  & = & m_g(\boldsymbol{x}_*) + \boldsymbol{K}_g (\boldsymbol{x}_*, \boldsymbol{x}_{1:N})[\boldsymbol{K}_g(\boldsymbol{x}_{1:N}, \boldsymbol{x}_{1:N}) \qquad\quad\nonumber\\
%&& \qquad\qquad\qquad\negmedspace{} + \boldsymbol{I}_N \otimes \boldsymbol{R}]^{-1}[\boldsymbol{y}_{1:N} - \boldsymbol{m}_g (\boldsymbol{x}_{1:N})], \IEEEyesnumber\IEEEyessubnumber\\
%\cov(\boldsymbol{y}_*) & = & k_g(\boldsymbol{x}_*, \boldsymbol{x}_*) - \boldsymbol{K}_g(\boldsymbol{x}_*, \boldsymbol{x}_{1:N})[\boldsymbol{K}_g(\boldsymbol{x}_{1:N}, \boldsymbol{x}_{1:N})  \qquad\quad\nonumber \\
%&& \qquad\qquad\qquad\qquad\negmedspace{} + \boldsymbol{I}_N  \otimes \boldsymbol{R}]^{-1}\boldsymbol{K}_g(\boldsymbol{x}_{1:N}, \boldsymbol{x}_*). \IEEEyessubnumber
%\end{IEEEeqnarray}

\vspace{-0.2\baselineskip}
\subsection{Learning of Transition Function $\boldsymbol{f}$}

\subsubsection{Pedestrian Dead Reckoning (PDR) for Control Input}
For indoor navigation, the control inputs $\boldsymbol{u}_{1:T}\triangleq\{\boldsymbol{u}_t\}_{t=1}^T$ of a SSM can be obtained by the widely-adopted PDR approach \cite{Chen15, Zou17} using the smartphone built-in IMUs such as the accelerometer, gyroscope, magnetometer, etc.~The PDR provides $\boldsymbol{u}_{1:T}$ for the corresponding state trajectory $\boldsymbol{x}_{0:T}$ through step detection, step length estimation and walking direction estimation,  which can be expressed as $\boldsymbol{u}_t = L_t [\sin(\psi_t), \cos(\psi_t)]^T, t = 1, 2, \ldots, T$, 
%
%\begin{equation}
%\boldsymbol{u}_t = L_t \begin{bmatrix}
%\sin(\psi_t) \\ \cos(\psi_t)
%\end{bmatrix}, \quad t = 1, 2, \ldots, T,
%\end{equation}
%
where $L_t$ and $\psi_t$ are respectively the step length and the walking direction of the pedestrian at time instance $t$.

Assume that we hold the smartphone with the front facing our walking direction. To detect steps, we first perform a rolling mean of the vertical acceleration readings of the linear acceleration sensor (effectively a composite sensor \cite{Android} based on accelerometer and gyroscope) to filter out the high-frequency noise, then find peaks (i.e., the local maxima) in the filtered data by comparing the neighboring values. %Fig.~\subref*{subfig:vertical acceleration} shows the vertical acceleration signal before and after the processing, along with the identified peaks. 
The walking direction $\psi_t$ could be estimated based on the rotation vector sensor output, which integrates the accelerometer, gyroscope, and magnetometer readings to compute the orientation of the smartphone \cite{Android}. Specifically, we compute the angle between the smartphone's pointing direction and the magnetic north pole based on the quaternion representation of the rotation vector. %Fig.~\subref*{subfig:azimuth} shows the estimated walking direction of a trajectory with two physical turns.
For simplicity, the step length estimate is fixed to a constant value $L_\text{const}$ throughout this paper.
%
%\begin{figure}[t]
%	\centering
%	\subfloat[Step Detection]{\includegraphics[width=0.4\columnwidth]{figure/accel}\label{subfig:vertical acceleration}}
%	\hfil
%	\subfloat[Walking Direction]{\includegraphics[width=0.4\columnwidth]{figure/azimuth}\label{subfig:azimuth}}
%	\caption{Step detection and walking direction estimation in PDR.}
%\end{figure}
%
The PDR determines the current position $\boldsymbol{x}_t$ by updating the previous position $\boldsymbol{x}_{t-1}$ based upon the current control input $\boldsymbol{u}_t$. Note that the control inputs $\boldsymbol{u}_{1:T}$ derived from the raw sensor measurements tend to provide biased estimates. Nevertheless, the empirical knowledge from PDR can be encoded as the mean function of the GP transition model, for instance, $m_f(\boldsymbol{x}_{t-1}, \boldsymbol{u}_t) = \boldsymbol{x}_{t-1} + \boldsymbol{u}_t$. 

%\vspace{-0.5\baselineskip}
\subsubsection{Variational Learning}
Unlike learning the measurement function $g$ at fixed calibration grids (like in fingerprinting), offline learning of the transition function $f$, requires the precise positions of a continuously walking pedestrian at discrete time instances, which are difficult to measure precisely with affordable time and workforce.~Therefore, in the learning phase we aim to find the smoothing distribution $p(\boldsymbol{x}_0, \boldsymbol{x}_1, \ldots, \boldsymbol{x}_T|\boldsymbol{y}_1, \boldsymbol{y}_2, \ldots, \boldsymbol{y}_T)$ of the latent states and meanwhile optimize the model hyperparameters of $f$, given the position estimates $\boldsymbol{y}_{1:T}$ and the control inputs $\boldsymbol{u}_{1:T}$. %Meanwhile, direct Bayesian learning of $f$ is intractable due to its complex structure. 
%To this end, we adopt the variational learning approach \cite{Frigola14a}, which approximates the posterior over latent variables in GPSSM with a variational distribution. The details are as follows.
% by minimizing the Kullback-Leibler (KL) divergence between the exact posterior and the approximating distribution. %Note that, the variational learning approach is used to learn \emph{only} $f$ since we have already learned the measurement function $g$. %This simplifies the optimization procedure and alleviates non-identifiabilities between $f$ and $g$.

As in \cite{Frigola14a}, the variational sparse GP framework \cite{Titsias09} is employed in order to deal with the unobserved inputs and outputs of $f$, as well as to reduce the computational complexity for learning $f$. To fake the position estimates, $M$ auxiliary inducing points $\boldsymbol{v}_{1:M}\triangleq\{\boldsymbol{v}_m\}_{m=1}^M$ are introduced, which are the values of $f$ evaluated at the inducing inputs $\boldsymbol{z}_{1:M}\triangleq\{\boldsymbol{z}_m\}_{m=1}^M$. %Therefore, $p(\boldsymbol{v}_{1:M}|\boldsymbol{z}_{1:M}, \boldsymbol{\theta}_f) = \mathcal{N}(\boldsymbol{v}_{1:M}|\boldsymbol{m}_f(\boldsymbol{z}_{1:M}), \boldsymbol{K}_f(\boldsymbol{z}_{1:M}, \boldsymbol{z}_{1:M}))$. 
The inducing points are jointly Gaussian with the latent function values $\boldsymbol{f}_{1:T}\triangleq\{\boldsymbol{f}_t\}_{t=1}^T$, since they are drawn from the same GP prior. Then the variational inference procedure \cite{Zhang18} is applied to circumvent the computationally intractable log-marginal likelihood of the GPSSM (cf. the LHS of (4)) by using, as replacement, its Evidence Lower BOund (ELBO), cf. the RHS of (4) below:
\begin{IEEEeqnarray}{l}
\log p(\boldsymbol{y}_{1:T})\!\!\geq\! \mathbb{E}_{q(\boldsymbol{x}_{0:T}, \boldsymbol{f}_{1:T}, \boldsymbol{v}_{1:M})}\!\!\left[\!\log\! \frac{p(\boldsymbol{y}_{1:T}, \!\boldsymbol{x}_{0:T}, \!\boldsymbol{f}_{1:T}, \! \boldsymbol{v}_{1:M})}{q(\boldsymbol{x}_{0:T},\! \boldsymbol{f}_{1:T},\! \boldsymbol{v}_{1:M})}\!\right]\!\!,\!\IEEEeqnarraynumspace
\end{IEEEeqnarray}
where the varaitional distribution $q(\boldsymbol{x}_{0:T}, \boldsymbol{f}_{1:T}, \boldsymbol{v}_{1:M})$ is introduced to approximate the true posterior $p(\boldsymbol{x}_{0:T}, \boldsymbol{f}_{1:T}, \boldsymbol{v}_{1:M}|\boldsymbol{y}_{1:T})$.
%Note that we omit the explicit conditioning on the inducing inputs $\boldsymbol{z}_{1:M}$ as well as hyperparameters $\boldsymbol{\theta}_f$ and $\boldsymbol{Q}$ for notational brevity. 
The jo\-int distribution of all latent variables is given by $p(\boldsymbol{y}_{1:T}, \boldsymbol{x}_{0:T}, \boldsymbol{f}_{1:T}, \allowbreak \boldsymbol{v}_{1:M}) =  p(\boldsymbol{f}_{1:T}|\boldsymbol{v}_{1:M})p(\boldsymbol{v}_{1:M})p(\boldsymbol{x}_0)\prod_{t=1}^T p(\boldsymbol{y}_t|\boldsymbol{x}_t)p(\boldsymbol{x}_t|\boldsymbol{f}_t)$, in which $p(\boldsymbol{f}_{1:T}|\boldsymbol{v}_{1:M})$ is the GP posterior evaluated at test inputs $\hat{\boldsymbol{x}}_{0:T-1} \allowbreak \triangleq \{\hat{\boldsymbol{x}}_t\}_{t=0}^{T-1}$ with the inducing points, %i.e.,
%
%\begin{IEEEeqnarray}{l}
%p(\boldsymbol{f}_{1:T}|\boldsymbol{v}_{1:M}) = \mathcal{N}(\boldsymbol{f}_{1:T} | \boldsymbol{m}_f(\hat{\boldsymbol{x}}_{0:T-1}) + \negmedspace {} \nonumber\\
%\boldsymbol{K}_f(\hat{\boldsymbol{x}}_{0:T-1}, \boldsymbol{z}_{1:M})\boldsymbol{K}_f(\boldsymbol{z}_{1:M}, \boldsymbol{z}_{1:M})^{-1}[\boldsymbol{v}_{1:M} - \boldsymbol{m}_f(\boldsymbol{z}_{1:M})],\nonumber\\
%\boldsymbol{K}_f(\hat{\boldsymbol{x}}_{0:T-1}, \hat{\boldsymbol{x}}_{0:T-1}) - \negmedspace {} \nonumber\\
%\boldsymbol{K}_f(\!\hat{\boldsymbol{x}}_{0:T-1}, \boldsymbol{z}_{1:M}\!)\boldsymbol{K}_f(\!\boldsymbol{z}_{1:M}, \boldsymbol{z}_{1:M}\!)^{-1}\boldsymbol{K}_f(\!\boldsymbol{z}_{1:M}, \hat{\boldsymbol{x}}_{0:T-1}\!)),\IEEEeqnarraynumspace\!
%\end{IEEEeqnarray}
%
and $p(\boldsymbol{y}_t|\boldsymbol{x}_t)$ is given in Section~\ref{subsubsec:GP Optimization} based on a learned measurement function $g$. The shorthand notation $\hat{\boldsymbol{x}}_{t-1} \triangleq \{\boldsymbol{x}_{t-1}, \boldsymbol{u}_{t}\}$ is used to denote an augmented input at time instance $t$. %Note that in $p(\boldsymbol{f}_{1:T}|\boldsymbol{v}_{1:M})$ the conditioning on $\hat{\boldsymbol{x}}_{0:T-1}$ is implicit since $\boldsymbol{f}_t$ is a function of $\hat{\boldsymbol{x}}_{t-1}$. 
%The ELBO serves as an objective function to be maximized for obtaining a point estimate of the model hyperparameters as well as the optimal set of inducing inputs. 
%It is well known \cite{Zhang18} that maximizing the ELBO is equivalent to minimizing the KL divergence between the intractable true posterior and its valid approximation.
%thereby obtaining the optimal posterior approximation over latent variables

In order to give a tractable ELBO, the approximated posterior $q(\boldsymbol{x}_{0:T}, \boldsymbol{f}_{1:T}, \boldsymbol{v}_{1:M})$ can be chosen, like in \cite{Frigola14a}, as $q(\boldsymbol{x}_{0:T}, \boldsymbol{f}_{1:T}, \allowbreak \boldsymbol{v}_{1:M}) = q(\boldsymbol{x}_{0:T}) p(\boldsymbol{f}_{1:T}|\boldsymbol{v}_{1:M}) q(\boldsymbol{v}_{1:M})$, 
%
%\begin{equation}
%q(\boldsymbol{x}_{0:T}, \boldsymbol{f}_{1:T}, \boldsymbol{v}_{1:M}) = q(\boldsymbol{x}_{0:T}) p(\boldsymbol{f}_{1:T}|\boldsymbol{v}_{1:M}) q(\boldsymbol{v}_{1:M}),\label{eq:posterior approximation}
%\end{equation}
%
where it is assumed that: 1) $q(\boldsymbol{\cdot})$ factorizes between $\boldsymbol{f}_{1:T}$ and $\boldsymbol{x}_{0:T}$ (a.k.a.\ the mean field approximation in variational inference literature), and 2) $\boldsymbol{f}_{1:T}$ is conditionally independent of $\boldsymbol{y}_{1:T}$ given the inducing points $\boldsymbol{v}_{1:M}$. Therefore, the ELBO can be formulated as a function of the posterior approximations $q(\boldsymbol{x}_{0:T})$ and $q(\boldsymbol{v}_{1:M})$, of the hyperparameters $\boldsymbol{\theta}_f, \boldsymbol{Q}$, and of the inducing inputs $\boldsymbol{z}_{1:M}$, like in \cite{Frigola15}, as
\begin{IEEEeqnarray}{rcl}
\IEEEeqnarraymulticol{3}{l}{\mathcal{L}(q(\boldsymbol{x}_{0:T}), q(\boldsymbol{v}_{1:M}), \boldsymbol{\theta}_f, \boldsymbol{Q}, \boldsymbol{z}_{1:M})} \nonumber\\
& = & -\mathrm{KL}(q(\boldsymbol{v}_{1:M})\|p(\boldsymbol{v}_{1:M})) + \mathcal{H}(q(\boldsymbol{x}_{0:T})) + \mathbb{E}_{q(\boldsymbol{x}_0)}[\log p(\boldsymbol{x}_0)] \nonumber\\
&& \negmedspace {} + \textstyle{\sum_{t=1}^{T}}\mathbb{E}_{q(\boldsymbol{x}_{t-1:t})q(\boldsymbol{v}_{1:M})}[\mathbb{E}_{p(\boldsymbol{f}_t|\boldsymbol{v}_{1:M})}[\log p(\boldsymbol{x}_t|\boldsymbol{f}_t)]] \nonumber\\
&& \negmedspace {} + \textstyle{\sum_{t=1}^{T}}\mathbb{E}_{q(\boldsymbol{x}_{t})}[\log p(\boldsymbol{y}_t|\boldsymbol{x}_t)],
\label{eq:ELBO}
\end{IEEEeqnarray}
where $\mathrm{KL}(\cdot\|\cdot)$ denotes the KL divergence between two distributions and $\mathcal{H}(\cdot)$ denotes the entropy of a distribution.

The optimal $q(\boldsymbol{v}_{1:M})$ and $q(\boldsymbol{x}_{0:T})$ (denoted as $q^*(\boldsymbol{v}_{1:M})$ and $q^*(\boldsymbol{x}_{0:T})$ in the sequel) are found by maximizing the ELBO in~\eqref{eq:ELBO} with fixed $\{\boldsymbol{\theta}_f, \boldsymbol{Q}, \boldsymbol{z}_{1:M}\}$, using calculus of variations. Consequently, $q^*(\boldsymbol{v}_{1:M})$ turns out to be Gaussian $\mathcal{N}(\boldsymbol{v}_{1:M}|\boldsymbol{\mu}, \boldsymbol{\Sigma})$, like in \cite{Frigola15}, with the natural parameters $\boldsymbol{\eta}_1 = \boldsymbol{K}_f(\boldsymbol{z}_{1:M}, \boldsymbol{z}_{1:M})^{-1}\boldsymbol{m}_f(\boldsymbol{z}_{1:M}) + \sum_{t=1}^{T} \mathbb{E}_{q(\boldsymbol{x}_{t-1:t})}[\boldsymbol{A}_{t-1}^T\boldsymbol{Q}^{-1}[\boldsymbol{x}_t - m_f(\hat{\boldsymbol{x}}_{t-1})]]$, and $\boldsymbol{\eta}_2 = -(1/2)\{ \allowbreak \boldsymbol{K}_f(\boldsymbol{z}_{1:M}, \boldsymbol{z}_{1:M})^{-1} +\sum_{t=1}^{T}\mathbb{E}_{q(\boldsymbol{x}_{t-1})}[\boldsymbol{A}_{t-1}^T\boldsymbol{Q}^{-1}\boldsymbol{A}_{t-1}]\}$,
%
%\begin{IEEEeqnarray}{rcl}
%\boldsymbol{\eta}_1 & = & \sum_{t=1}^{T} \mathbb{E}_{q(\boldsymbol{x}_{t-1:t})}[\boldsymbol{A}_{t-1}^T\boldsymbol{Q}^{-1}[\boldsymbol{x}_t - m_f(\hat{\boldsymbol{x}}_{t-1})]] \nonumber \\
%&& \negmedspace {} + \boldsymbol{K}_f(\boldsymbol{z}_{1:M}, \boldsymbol{z}_{1:M})^{-1}\boldsymbol{m}_f(\boldsymbol{z}_{1:M}),\IEEEyesnumber\IEEEyessubnumber\\
%\boldsymbol{\eta}_2 & = & -\frac{1}{2}\!\bigg[\!\boldsymbol{K}_f(\boldsymbol{z}_{1:M}, \boldsymbol{z}_{1:M})\!^{-1}\!\! +\!\!\sum_{t=1}^{T}\mathbb{E}_{q(\boldsymbol{x}_{t-1})}[\boldsymbol{A}_{t-1}^T\boldsymbol{Q}^{-1}\!\!\boldsymbol{A}_{t-1}]\!\bigg]\!,\,\IEEEeqnarraynumspace\IEEEyessubnumber
%\end{IEEEeqnarray}
%                                      
where $\boldsymbol{A}_{t-1} \triangleq \boldsymbol{K}_f(\hat{\boldsymbol{x}}_{t-1},\boldsymbol{z}_{1:M})\boldsymbol{K}_f(\boldsymbol{z}_{1:M}, \boldsymbol{z}_{1:M})^{-1}$. The mean vector and covariance matrix of $q^*(\boldsymbol{v}_{1:M})$ can be computed as $\boldsymbol{\mu} = \boldsymbol{\Sigma}\boldsymbol{\eta}_1$ and $\boldsymbol{\Sigma} = (-2\boldsymbol{\eta}_2)^{-1}$. This leads to
\begin{IEEEeqnarray}{l}
q^*(\boldsymbol{x}_{0:T}) \propto p(\boldsymbol{x}_0)\big[\textstyle{\prod_{t=1}^{T}}\mathcal{N}(\boldsymbol{x}_t|m_f(\hat{\boldsymbol{x}}_{t-1}) + \boldsymbol{A}_{t-1}\boldsymbol{\mu}, \boldsymbol{Q})\big]\nonumber\\
\IEEEeqnarraymulticol{1}{r}{
\!\!\!\qquad\negmedspace {}\cdot\textstyle{\prod_{t=1}^{T}}p(\boldsymbol{y}_t|\boldsymbol{x}_t)\exp\big[\!-\!\frac{1}{2}\!\tr[\boldsymbol{Q}^{-1}(\boldsymbol{B}_{t-1} + \boldsymbol{A}_{t-1}\boldsymbol{\Sigma}\boldsymbol{A}_{t-1}^T)]\big]\!,\!\!\!\!\IEEEeqnarraynumspace}
\end{IEEEeqnarray}
where $\boldsymbol{B}_{t-1} \triangleq \boldsymbol{K}_f(\hat{\boldsymbol{x}}_{t-1},\hat{\boldsymbol{x}}_{t-1}) - \boldsymbol{K}_f(\hat{\boldsymbol{x}}_{t-1},\boldsymbol{z}_{1:M})\boldsymbol{K}_f(\boldsymbol{z}_{1:M}, \allowbreak \boldsymbol{z}_{1:M})^{-1}\boldsymbol{K}_f(\boldsymbol{z}_{1:M}, \hat{\boldsymbol{x}}_{t-1})$. This optimal form can be interpreted as the smoothing distribution of an auxiliary parametric SSM \cite{Frigola15}. Smoothing in such a nonlinear Markovian SSM can be done through, for instance, a particle smoother \cite{Doucet09}. 

When fixing $q(\boldsymbol{v}_{1:M})$ and $q(\boldsymbol{x}_{0:T})$, the model hyperparameters $\boldsymbol{\theta}_f$ and $\boldsymbol{Q}$, together with the inducing inputs $\boldsymbol{z}_{1:M}$ can be optimized by maximizing the ELBO using gradient-based methods. The gradient of ELBO in~\eqref{eq:ELBO} w.r.t.\ $\boldsymbol{\theta} \triangleq \{\boldsymbol{\theta}_f, \boldsymbol{Q}, \boldsymbol{z}_{1:M}\}$ is computed as
\begin{IEEEeqnarray}{rl}
\textstyle\frac{\partial\mathcal{L}}{\partial\boldsymbol{\theta}} & \textstyle= \mathbb{E}_{q(\boldsymbol{v}_{1:M})}[\frac{\partial}{\partial\boldsymbol{\theta}}\log p(\boldsymbol{v}_{1:M})] \nonumber\\
& \textstyle\!+ \! \sum_{t=1}^{T}\mathbb{E}_{q(\boldsymbol{x}_{t-1:t})}[\frac{\partial}{\partial\boldsymbol{\theta}} \log \mathcal{N}(\boldsymbol{x}_t|m_f(\hat{\boldsymbol{x}}_{t-1}) + \boldsymbol{A}_{t-1}\boldsymbol{\mu}, \boldsymbol{Q})] \nonumber\\
& \textstyle\! + \! \sum_{t=1}^{T}\mathbb{E}_{q(\boldsymbol{x}_{t-1})}[-\frac{1}{2}\frac{\partial}{\partial\boldsymbol{\theta}}\!\tr[\boldsymbol{Q}^{-1}(\boldsymbol{B}_{t-1} \!+\! \boldsymbol{A}_{t-1}\boldsymbol{\Sigma}\boldsymbol{A}_{t-1}^T)]].\IEEEeqnarraynumspace\!
\end{IEEEeqnarray}

In summary, like in \cite{Frigola15}, the ELBO~\eqref{eq:ELBO} is maximized by alternately 1) sampling from the smoothing distribution $q^*(\boldsymbol{x}_{0:T})$; 2) updating the natural parameters of $q^*(\boldsymbol{v}_{1:M})$; and 3) taking gradient steps in terms of $\{\boldsymbol{\theta}_f, \boldsymbol{Q}, \boldsymbol{z}_{1:M}\}$. %The optimal distribution of the initial state is simply $q^*(\boldsymbol{x}_0)$.
The above presented formulation and optimization procedure w.r.t.\ a single trajectory can be generalized to deal with multiple trajectories, since the ELBO~\eqref{eq:ELBO} can be factorized across independent time series. Stochastic optimization \cite{Hensman13} can therefore be used to maximize the ELBO, thereby alleviating computational burden when learning a large number of trajectories. 
%Specifically, in every iteration, one randomly selects a trajectory (a.k.a.\ mini-batch) and samples from the smoothing distribution of its latent states to obtain stochastic estimates of $q^*(\boldsymbol{v}_{1:M})$ and $\frac{\partial\mathcal{L}}{\partial\boldsymbol{\theta}}$, which depend on contributions from all latent states of available trajectories.

%The variational learning approach gives a tractable approximate posterior over $f$ \cite{Frigola15}. The approximate posterior distribution of the latent state $\boldsymbol{x}_{*+1}$ at test point $\hat{\boldsymbol{x}}_*$ is given by
%
%\begin{IEEEeqnarray}{l}
%p(\boldsymbol{x}_{*+1}|\hat{\boldsymbol{x}}_*, \boldsymbol{y}_{1:T}) \nonumber \\
%\approx \mathcal{N}(\boldsymbol{x}_{*+1}|m_f(\hat{\boldsymbol{x}}_*) + \boldsymbol{A}_*\boldsymbol{\mu}, \boldsymbol{B}_* + \boldsymbol{A}_*\boldsymbol{\Sigma}\boldsymbol{A}_*^T + \boldsymbol{Q}).
%\end{IEEEeqnarray}

\begin{figure*}[t]
	\centering
	\subfloat[WiFi Localization Only (Static Estimates)]{\includegraphics[width=0.58\columnwidth]{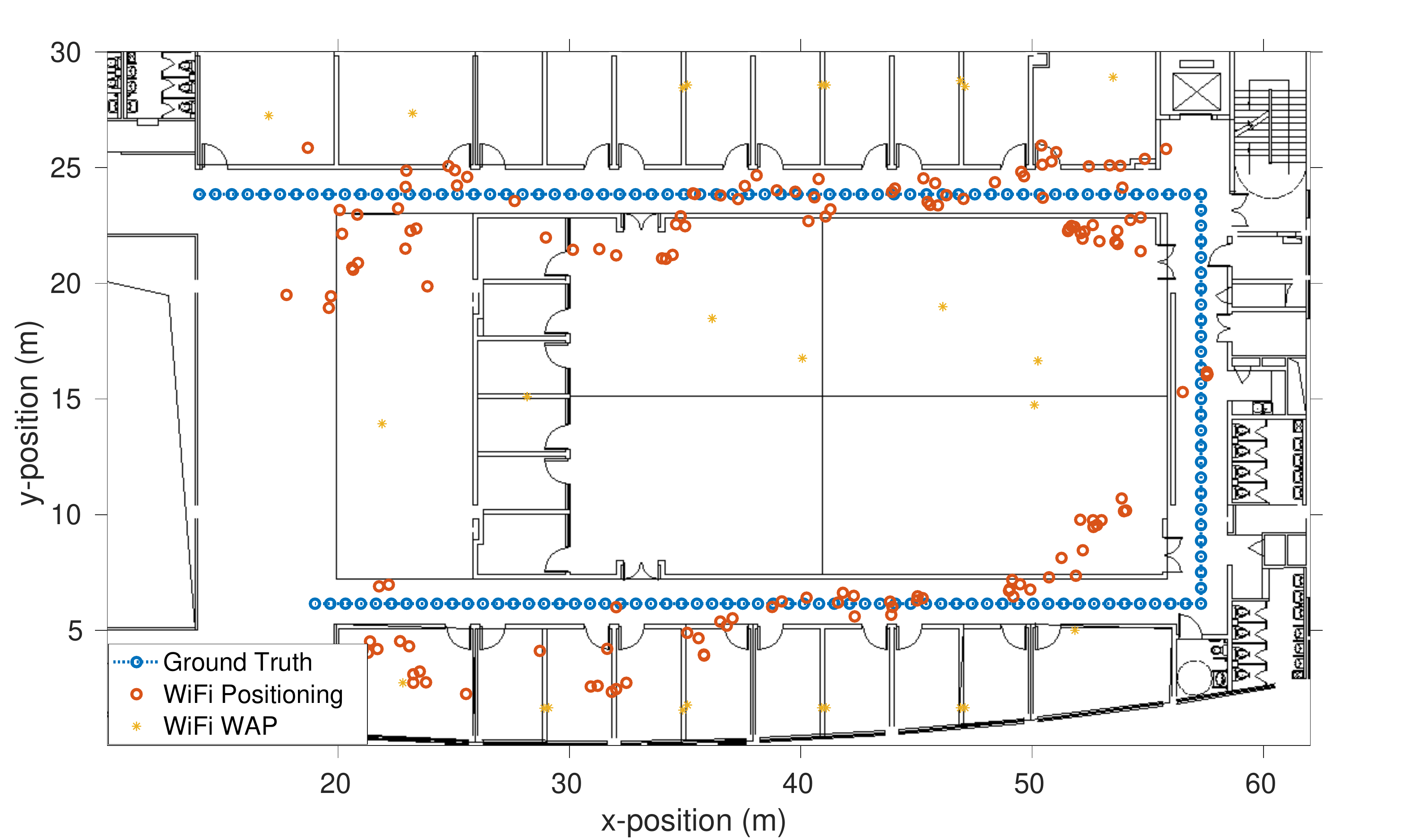}}
	\hfil
	\subfloat[PDR Only]{\includegraphics[width=0.58\columnwidth]{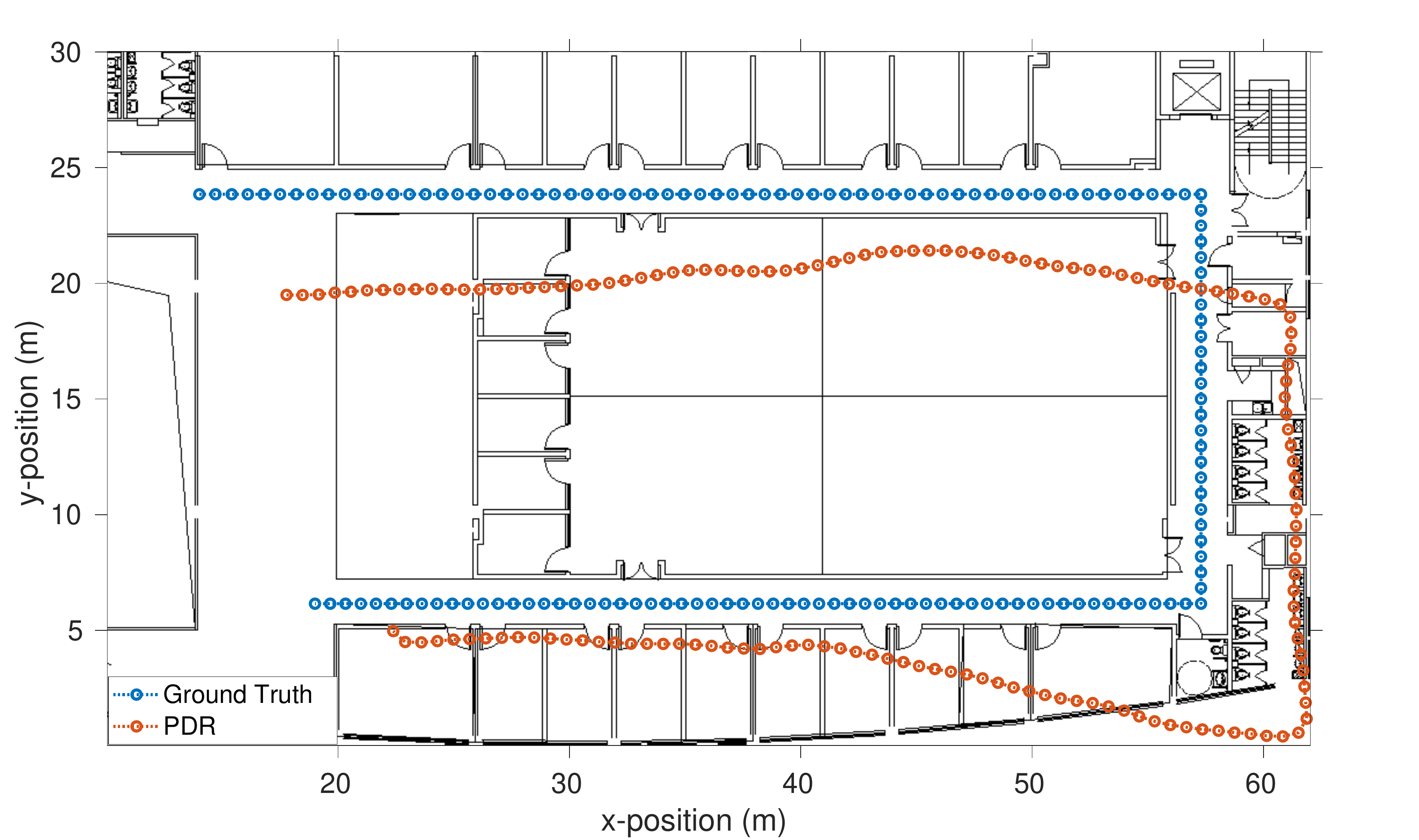}}
	\hfil
	\subfloat[LGSSM]{\includegraphics[width=0.58\columnwidth]{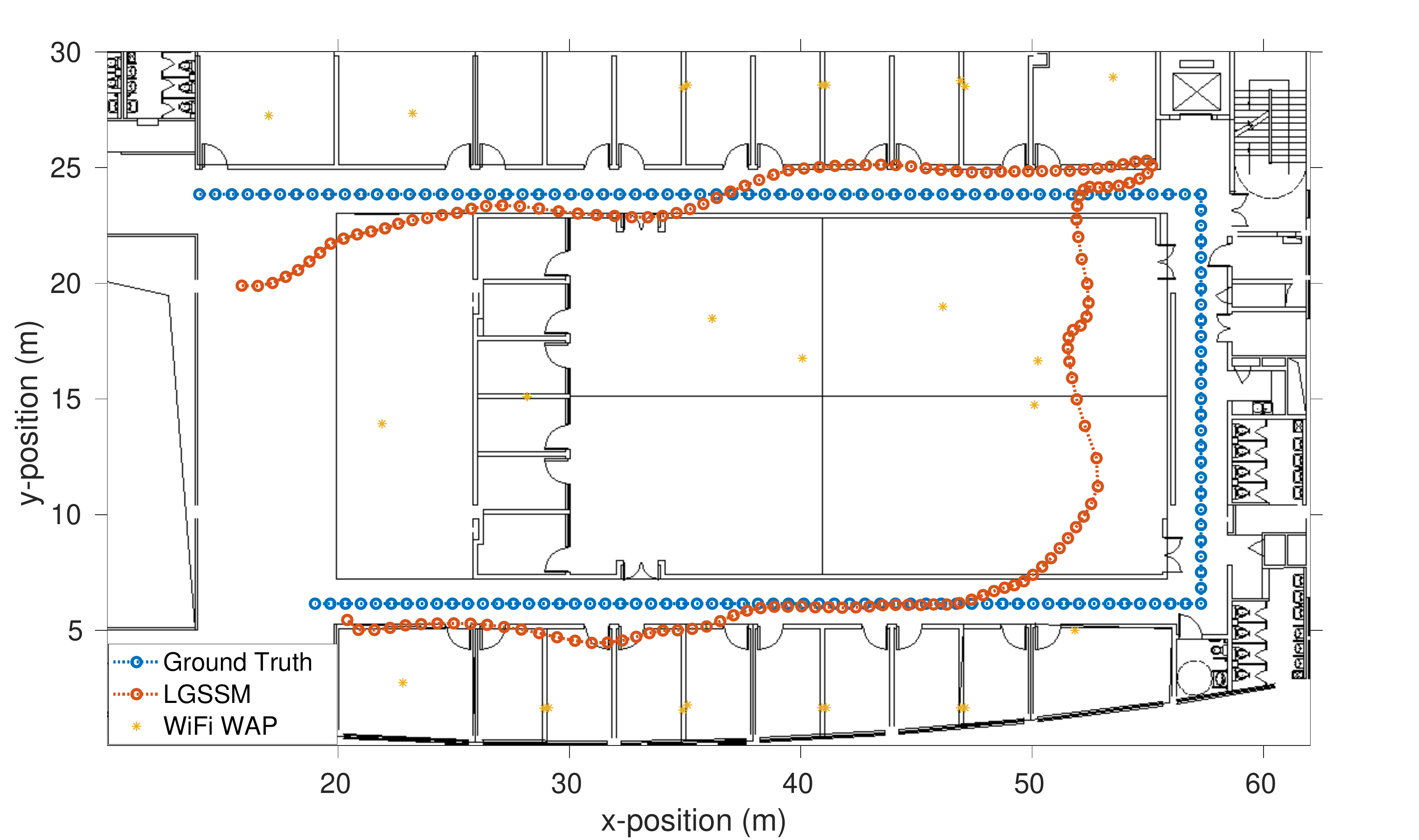}\label{subfig:LGSSM}}
	
	\vspace{-0.8\baselineskip}
	\subfloat[GPSSM with $1$ Trajectory Data]{\includegraphics[width=0.58\columnwidth]{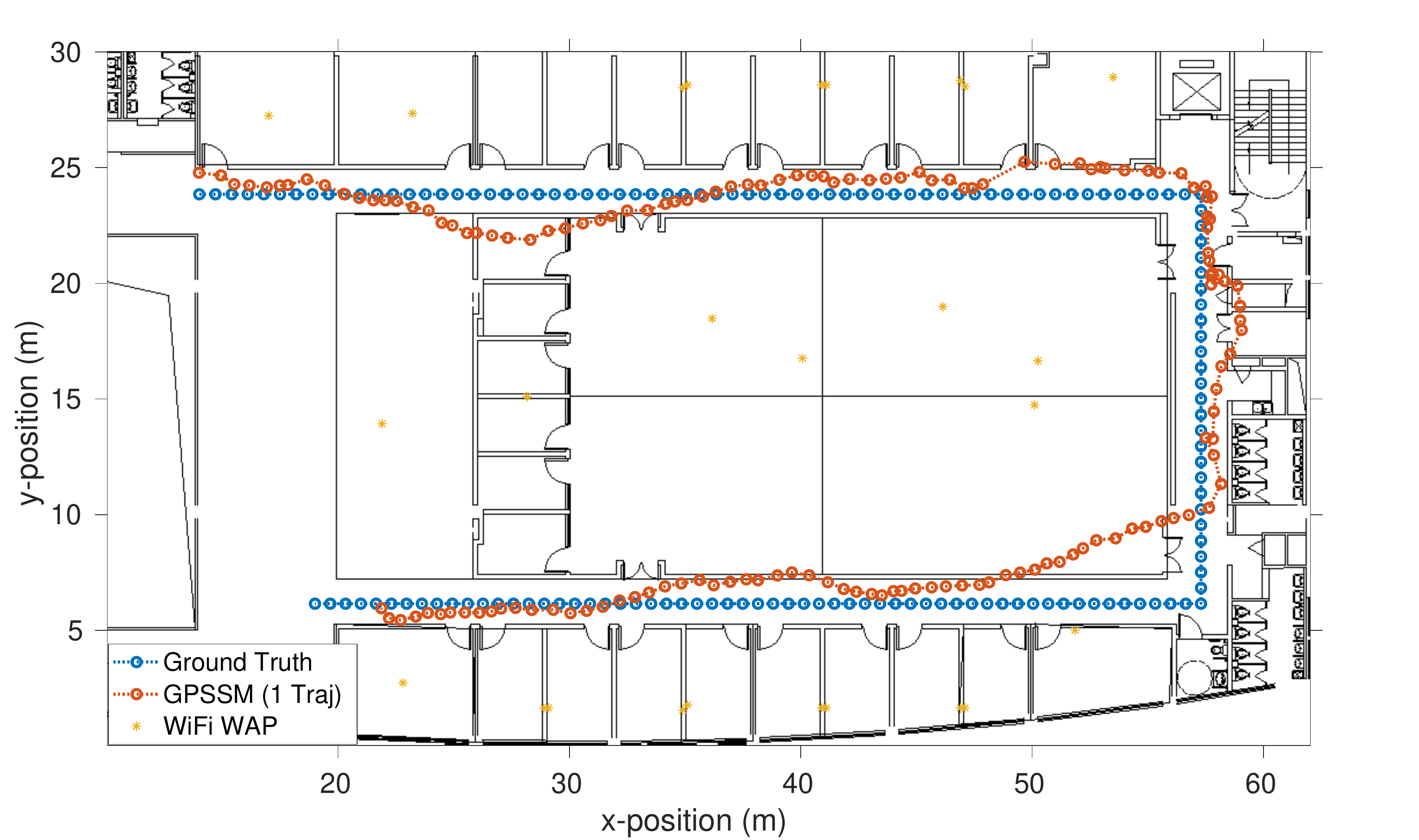}\label{subfig:GPSSM (1 Traj)}}
	\hfil
	\subfloat[GPSSM with $3$ Trajectory Data]{\includegraphics[width=0.58\columnwidth]{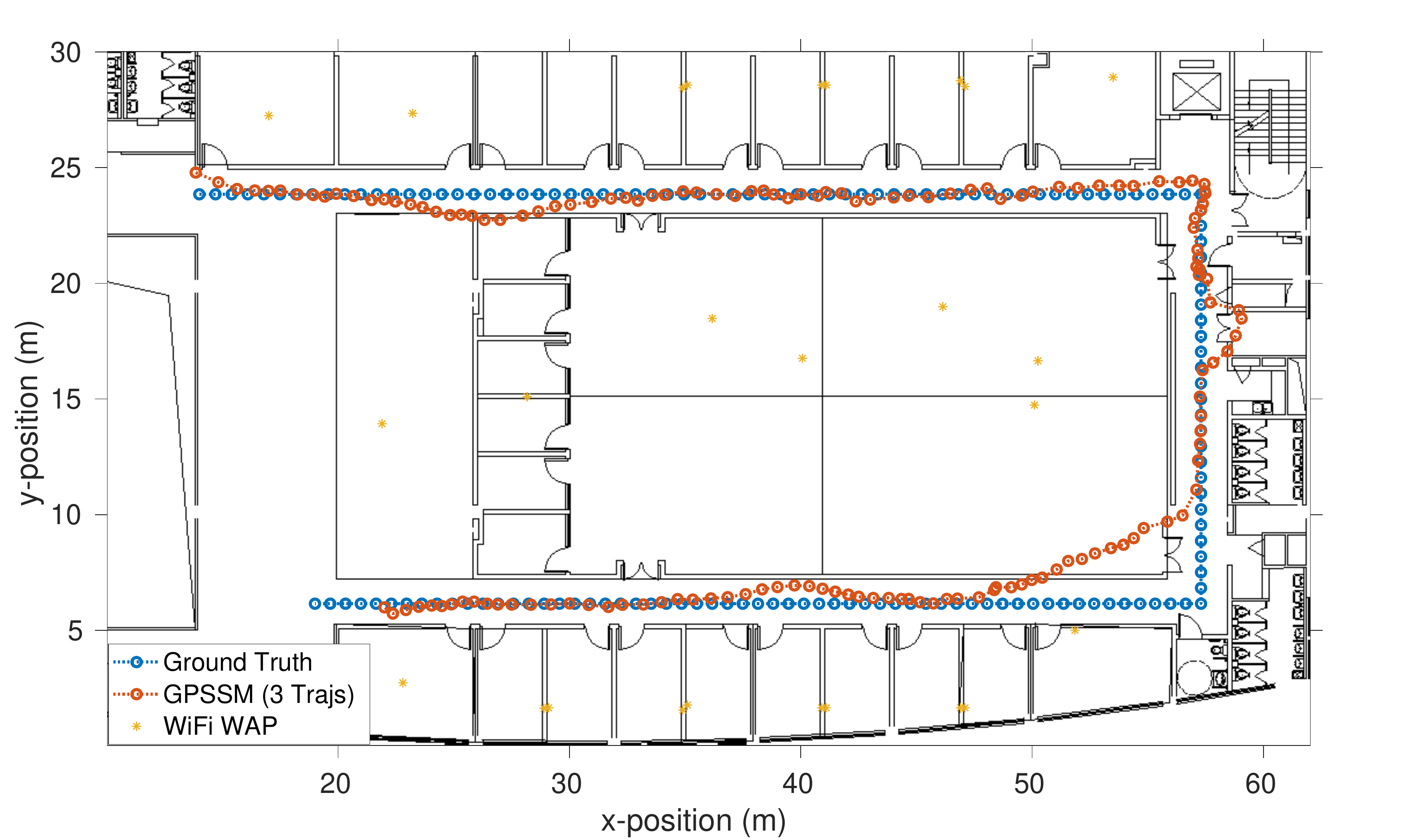}}
	\hfil
	\subfloat[GPSSM with $5$ Trajectory Data]{\includegraphics[width=0.58\columnwidth]{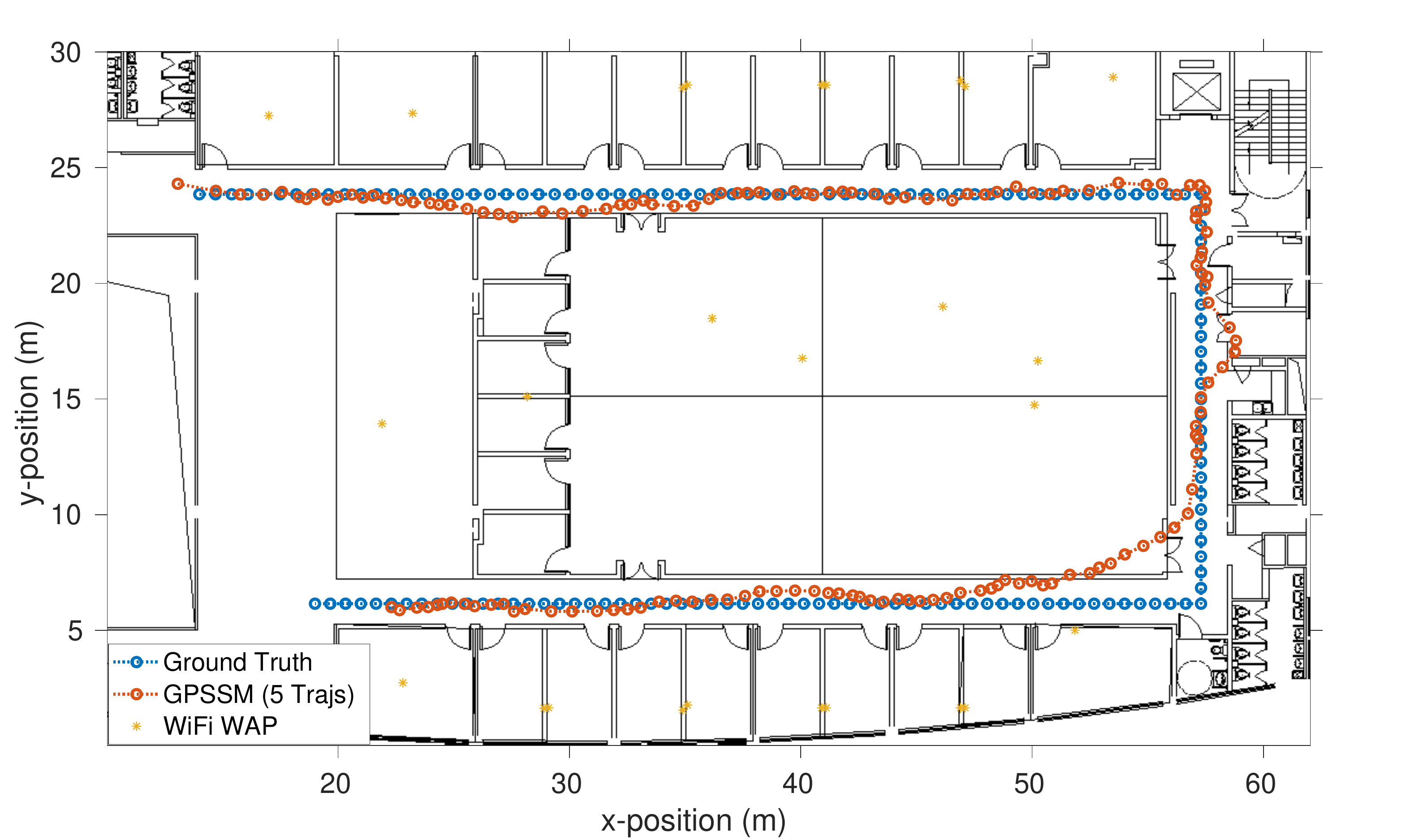}\label{subfig:GPSSM (5 Traj)}}
	\caption{Layout of the WiFi APs (yellow stars), the ground-truth trajectory (blue dots), and the recovered trajectories (orange dots).}
	\label{fig:trajectories}
	\vspace{-0.6\baselineskip}
\end{figure*}

%\vspace{-0.5\baselineskip}
\subsubsection{Practical Selection of Inducing Inputs}
Ideally, the inducing inputs $\boldsymbol{z}_{1:M}$ in the GPSSM should be optimized jointly with the model hyperparameters. However, direct optimization of the inducing inputs tends to find under-fitting solutions as $M$ becomes large \cite{Bauer16}. In practice, we prefer to select several representative trajectories among all training trajectories based on certain criteria, and choose $\boldsymbol{z}_{1:M}$ to be the latent states sampled from the smoothing distributions of the selected trajectories, concatenated with the corresponding control inputs. The aforementioned ELBO can be used as a suitable criterion to maximize for selecting these representative trajectories \cite{Titsias09}. Tuning the number of inducing inputs balances the approximation accuracy and the computational cost.

%\vspace{-0.5\baselineskip}
\section{Experimental Results}
\label{sec:Experiments}
To evaluate the performance of the proposed indoor navigation system, we conduct various experiments in a $1600$ m$^2$ office environment at The Chinese University of Hong Kong, Shenzhen. We aim to recover a ``U''-shape walking trajectory of a pedestrian holding a smartphone. The geographical layout of the $26$ WiFi APs in the deployment area is shown in Fig.~\ref{fig:trajectories}. We developed a mobile application on a HUAWEI smartphone running Android 7.0 operating system to collect WiFi RSS measurements as well as the other sensory data. A sampling rate of $100$~Hz is specified for the IMU sensors. The collected data are transmitted via wireless links to a computing server for further processing. 
%In the future, all the processing and computations will be designed to run on the mobile. 

To learn the measurement function $g$ with a GP model, a total number of $N = 2059$ ground-truth positions and their corresponding WiFi RSS measurements were recorded across the whole area. We adopt a threshold value for RSS collection according to \cite{Yin17a}. To learn the transition function $f$ with a GP model, we recorded the WiFi RSS measurements as well as the PDR control inputs while walking in the area. We repetitively record up to $5$ trajectories (loops) along the same predefined path, and recover the first state trajectory that is tracked for $147$ steps. %The initial guess of the first latent state of each trajectory is set to be the first measurement of the corresponding trajectory. 
For both $f$ and $g$, we chose the standard squared exponential (SE) kernel with automatic relevance determination (ARD) \cite{RW06} in the GP models. For simplicity, we model each output dimension of the GP independently. 
%The first trajectory is selected as the representative trajectory to provide the inducing inputs. 
We use the conjugate gradient method to optimize the GP hyperparameters of the measurement function $g$, while using the stochastic gradient descent to optimize the GP hyperparameters of the transition function $f$. For comparison, we learn a linear Gaussian state-space model (short as LGSSM) using the EM algorithm with the measurements and control inputs of the first trajectory.

\begin{table}[t]
	\vspace{-0.2\baselineskip}
	\caption{Navigation Accuracy in Terms of MAE.}
	\label{table:MAE}
	\centering
	\scriptsize
	\begin{tabular}{ccc}
		\toprule
		WiFi Localization Only & PDR Only & LGSSM \\
		\midrule
		$3.55$ m & $5.34$ m & $2.72$ m  \\
		\bottomrule
		\toprule
		GPSSM ($1$ Traj) & GPSSM ($3$ Trajs) & GPSSM ($5$ Trajs) \\
		\midrule
		$2.29$ m & $2.16$ m & $2.11$ m \\
		\bottomrule
	\end{tabular}
	\vspace{-1.4\baselineskip}
\end{table}

In Fig.~\ref{fig:trajectories}, we show the state trajectories recovered by different models. The corresponding navigation mean-absolute-error (MAE) is reported in Table~\ref{table:MAE}. Specifically, the WiFi localization gives unsatisfactory position estimates in the trajectory segment around the two sharp turns. This is due to the lack of APs deployed in that area and the unreliable RSS values received from far-away APs. PDR recovers a drifted state trajectory compared to the ground-truth. Since PDR only provides relative information, we use the WiFi localization estimate as the initial position for PDR. Clearly, both the WiFi localization and PDR are unsatisfactory. However, fusing WiFi localization and PDR in the LGSSM or GPSSM takes the advantages of the individual techniques, and achieves higher navigation accuracy. The parametric LGSSM models the underlying dynamics and the measurement mapping poorly, therefore does not provide satisfactory trajectory recovery. It is obvious from Fig.~\subref*{subfig:LGSSM} that the LGSSM underestimates the measurement noise and is largely misled by the poor WiFi localization estimates. Contrarily, the non-parametric GPSSM empowered by the proposed learning procedure is able to recover the state trajectories closest to the ground-truth. Fig.~\subref*{subfig:GPSSM (1 Traj)}--\subref*{subfig:GPSSM (5 Traj)} show that the GPSSM keeps improving the estimation quality when learning over more training data. The outstanding modeling capacity of the GPSSM is exploited by feeding the model with large datasets, which contains comprehensive information about the indoor environment as well as pedestrian's motion patterns. More importantly, the feasibility of the proposed GPSSM learning procedure  is proved, which circumvents the non-identifiability issue of the previous learning procedures, reduces optimization complexity, and makes practical implementation of the GPSSM favorable for indoor navigation. 

\section{Conclusion}
\label{sec:Conclusion}
In this paper, we applied the Gaussian process state-space model (GPSSM) for indoor navigation based on the proposed practical learning procedure that fuses WiFi RSS and readings from smartphone built-in IMU sensors. The non-parametric nature of GPSSM makes it a powerful modeling tool for complex behaviors in real-world indoor navigation. Experiments in a real office environment demonstrate the superior performance of GPSSM over the classical parametric SSMs in recovering the state trajectory from noisy measurements. 
%As future work, we plan to collect more trajectory data and conduct all processings and computations on the mobile phone. 

% The IEEEbib.bst bibliography style file from IEEE produces unsorted bibliography list.
% -------------------------------------------------------------------------
\newpage
%\IEEEtriggeratref{13}
\bibliographystyle{IEEEtran}
\bibliography{reference}

\end{document}